\documentclass[prl,showpacs]{revtex4}
\usepackage{graphicx}
\usepackage{dcolumn}
\usepackage{bm}
\usepackage{amssymb}
\usepackage{amsmath}

\begin{document}

\title{Theory of the Laser Wake-Field Accelerator Revisited: \\
Wake Overtaking, Localized Spectrum and Ponderomotive Acceleration}

\author{T. Esirkepov}
\altaffiliation[Also at]{Moscow Institute of Physics and Technology, Institutskij pereulok 9, Dolgoprudnyi, 141700 Russia.}
\affiliation{Kansai Research Establishment, JAERI, Kizu, Kyoto, 619-0215 Japan}

\author{S. V. Bulanov}
\altaffiliation[Also at]{General Physics Institute RAS, Vavilov street 38, Moscow, 119991 Russia.}
\affiliation{Kansai Research Establishment, JAERI, Kizu, Kyoto, 619-0215 Japan}

\author{M. Yamagiwa}
\author{T. Tajima}
\affiliation{Kansai Research Establishment, JAERI, Kizu, Kyoto, 619-0215 Japan}

\date{April, 2005}

\begin{abstract}
The electron and positron acceleration
in the first cycle of a laser-driven wakefield
is investigated.
Separatrices between different types of the particle motion
(confined, reflected by the wakefield or ponderomotive potential and transient)
are demonstrated.
The ponderomotive acceleration is negligible for electrons
but is substantial for positrons.
An electron bunch, injected as quasi-monoenergetic,
acquires a localized energy spectrum with a cut-off at the maximum energy.
\end{abstract}



\maketitle


Laser-driven charged particle acceleration
is an attractive alternative
to cyclic accelerators and linacs,
promising to provide
much greater acceleration rate
with a much more compact facility.
At the dawn of the laser technology
the ``optical maser'' was suggested 
in Ref. \cite{Shimoda}
to accelerate electrons.
The long-living strong Langmuir wave (wakefield),
left in the wake of a short intense laser pulse
in a low-density collisionless plasma,
accelerates duly injected electrons in
the Laser Wake-Field Accelerator (LWFA) concept
introduced 
in Ref. \cite{Tajima-Dawson}.
For efficient acceleration of charged particles, the laser pulse must be relativistically strong,
i. e., its amplitude $a_0 = e E_0/m_e \omega c>1$.
To provide electrons
one must use an externally preaccelerated electron bunch
or exploit the effect of self-injection
due to a longitudinal Langmuir wave-break \cite{break}
or/and a transverse wave-break \cite{Bulanov-TWB},
which dominates when the laser pulse waist
becomes comparable with or less than the wakefield wavelength,
e. g., due to the pulse self-focusing.
The injection of the preaccelerated electron bunch
co-propagating with the laser pulse was considered in Ref. \cite{KHACH}.
The wave-breaking occurs when the displacement
of the plasma electron fluid moving in the wave
becomes equal to or larger than the wakefield wavelength
$\lambda_{\rm wf} = 4\pi\sqrt{2}\, c\, a_0 /\omega_{pe}$,
where $\omega_{pe}=\sqrt{4\pi n_e e^2/m_e}$ is the Langmuir frequency.

In recent experiments
localized energy spectra of electrons accelerated
up to 170 MeV were demonstrated \cite{RAL,Recent-exp}.
The indications were given that the laser pulse
undergo a self-focusing, and the wave-breaking
(both the longitudinal and transverse)
occures in the first cycle of the wakefield
with the electron self-injection into the acceleration phase.
In Ref. \cite{RAL}
the localized electron energy spectrum formation
was attributed to 
electrons accelerated in the wakefield first cycle,
overtaking the laser pulse.
The so-called ``ponderomotive electron acceleration'',
suggested in Ref. \cite{G-M} and analyzed in Refs. \cite{PA},
describes the charged particle motion at the laser pulse front,
thus it is in effect also in the wakefield first cycle.
The positron acceleration
by a long electromagnetic wave in underdense plasmas
was considered in Ref. \cite{POSITRONS}.

In this Letter we revisit
theory of the Laser Wake-Field Accelerator
and examine the acceleration of charged particles
in the {\it first} cycle of the wakefield.
The electron energy spectrum
is calculated in a general case of nonoptimal injection.
The role of the ponderomotive acceleration is discussed
in the case of electrons and positrons.

%
%
In the framework of classical electrodynamics
the one-dimensional motion of a particle
with charge $-e$ and mass $m_e$
in the laser pulse and wakefield
is described by the Hamiltonian \cite{LL}
\begin{equation} \label{Hamiltonian}
\mathcal{H} = \sqrt{
m_e^2 c^4 + c^2 P_\parallel^2 +
\bigl( c P_\perp + e A_\perp (X) \bigr)^2
}
- e \varphi (X)
\, ,
\end{equation}
where $X = x - v_g t$, $x$ is the particle coordinate,
$v_g$ is the group velocity of the laser pulse (equal to the wakefield
phase velocity), $0<v_g<c$;
$P_\parallel$ and $P_\perp$ are the longitudinal and transverse components of
the generalized momentum,
$A_\perp$ is the laser pulse vector-potential,
$\varphi$ is the the wakefield potential.
In general, $P_\perp$ and $A_\perp$ have two components
(along $y$ and $z$).
Here we neglect the dispersion of the the laser pulse,
assuming that the laser pulse field depends
on time and coordinate as $A_\perp(x-v_g t)$.
The Hamiltonian (\ref{Hamiltonian}) admits a Lie group with generators
$v_g \partial_x + \partial_t$,
$\partial_y$, $\partial_z$.
Correspondingly, the Noether theorem
implies the motion integrals:
\begin{equation} \label{integrals}
\mathcal{H} - v_g P_\parallel = m_e c^2 h_0
\, ,\quad
P_\perp = P_{\perp 0}
\, ,
\end{equation}
where $h_0$ and $P_{\perp 0}$ are constants
of the particle initial momentum.
We introduce dimensionless variables
\begin{eqnarray}
\nonumber
\beta_{\rm ph} = v_g/c
\, ,
\Phi(X) = e\varphi(X)/m_e c^2
\, ,
p_x = P_\parallel/m_e c
\, ,
\\
a(X) = P_{\perp 0}/m_e c + e A_\perp(X)/m_e c^2
\, .
\label{subst}
\end{eqnarray}
We note that the variable $a(X)$
represents both the particle
initial transverse generalized momentum
and the laser pulse vector-potential.
In the case of $P_{\perp 0}=0$
it is just the laser dimensionless amplitude.
The first integral in (\ref{integrals})
in terms of new variables (\ref{subst}) gives
the equation
\begin{equation} \label{h-hamilt}
h(X,p_x) \stackrel{\text{def}}{=}
\sqrt{1+p_x^2+a^2(X)} - \Phi(X) - \beta_{\rm ph} p_x = h_0
\, .
\end{equation}
Its solution for $\beta_{\rm ph}<1$ is written as
\begin{eqnarray}
p_x &=&
\gamma_{\rm ph}^2
\Bigl\{
	\beta_{\rm ph} (\Phi(X) + h_0)
	\pm
\bigl[
(\Phi(X) + h_0)^2 -
\bigr.
\Bigr.
\nonumber
\\
&&
\Bigl.
\bigl.
\gamma_{\rm ph}^{-2} \left( 1+a^2(X) \right)
\bigr]^{1/2}
\Bigr\}
\label{p-sol}
\, ,
\end{eqnarray}
where $\gamma_{\rm ph} = (1-\beta_{\rm ph}^2)^{-1/2}$;
the sign `$+$' is for $X$ increasing with time
and `$-$' is for $X$ decreasing with time.
The particle moving from $X_0$ to $X$
with monotonically increasing $X(t)$
acquires the net kinetic energy
\begin{eqnarray}
{\cal E} &=&
{\cal E}_0 +
\gamma_{\rm ph}^2
\Bigl\{
	\Delta\Phi + \chi_0
	+ 
	\beta_{\rm ph}
\bigl[
(\Delta\Phi + \chi_0)^2 -
\bigr.
\Bigr.
\nonumber
\\
&&
\Bigl.
\bigl.
\gamma_{\rm ph}^{-2} \left( 1+a^2(X) \right)
\bigr]^{1/2}
\Bigr\}
-\chi_0 - \beta_{\rm ph} p_{x0}
\label{energy}
\, ,
\end{eqnarray}
where $\Delta\Phi = \Phi(X)-\Phi(X_0)$,
${\cal E}_0 = \sqrt{1+p_{x0}^2+a^2(X_0)} - 1$,
$\chi_0 = {\cal E}_0 + 1 - \beta_{\rm ph} p_{x0}$
and $p_{x0} = p_x(X_0)$.

%
%
%
%

\begin{figure}
\includegraphics{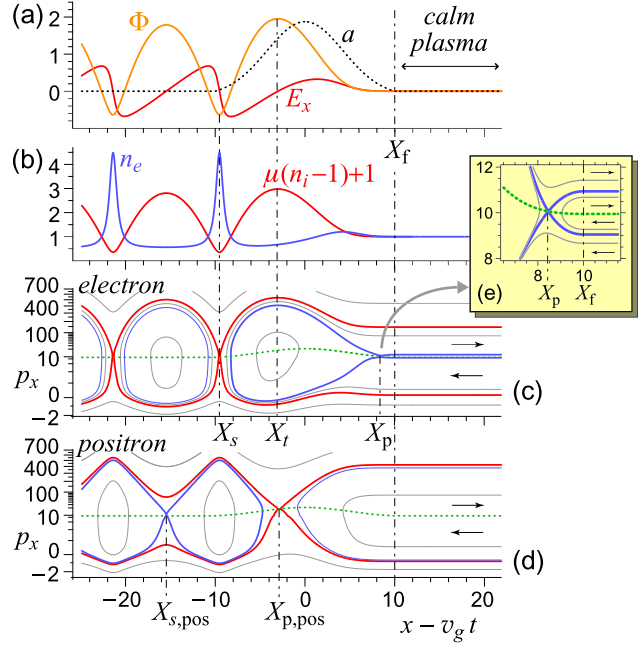}
\caption{\label{fig:phase}
(color).
The wakefield excited by the laser pulse (a);
electron and scaled ion density (b).
Phase portrait for the electron (c) and positron (d);
the electron ``ponderomotive'' basin close-up (e);
thick solid line for separatrices,
thin solid line for other orbits,
thick dotted line for
$p_{xs}(X) = \beta_{\rm ph} \gamma_{\rm ph} \sqrt{1+a^2(X)}$.
}
\end{figure}

To exemplify the general property of the
system with Hamiltonian $h(X,p_x)$,
we show its phase portrait
in Fig. \ref{fig:phase}(c), (e)
for the electron with $P_{\perp 0}=0$
in the case when the circularly polarized quasi-Gaussian laser pulse
with amplitude
$a(X)=\big\{ a_0 \left( \exp(-4\ln(2)X^2/l_p^2) - 1/16 \right) $
	at $|X| \leqslant l_p$, 0 at $|X|>l_p \big\} $,
$a_0=2$, FWHM size $l_p=10$ wavelengths,
propagates in an ideal Hydrogen plasma
with density $n_e = 0.01 n_{\rm cr}$,
and excites a wakefield,
whose potential is described by the Poisson equation, \cite{Phi-Eq},
\begin{eqnarray}
\Phi^{\prime\prime} &\!\!\!\! =\!\! &
k_p^2
\gamma_{\rm ph}^{3}
\beta_{\rm ph}
\!
\left\{
(1+\Phi)
	\bigl[\gamma_{\rm ph}^{2}(1+\Phi)^2 \! - 1 \! - a^2(X)\bigr]^{-1/2}
-
\right.
\nonumber
\\
&&
\left.
(\mu-\Phi)
	\bigl[\gamma_{\rm ph}^{2}(\mu - \Phi)^2 \! - \mu^2 \! - a^2(X)\bigr]^{-1/2}
\right\}
\, ,
\label{eq:phi}
\end{eqnarray}
where the prime denotes differentiation with respect to the $X$ coordinate,
$k_p = \omega_{pe}/c$ and
$\mu = m_i/m_e = 1836$ is the ion-to-electron mass ratio.
The potential $\Phi$, corresponding to the longitudinal electric field,
and the electron and ion densities as well as the laser pulse
envelope are shown in Fig. \ref{fig:phase}(a), (b).
We consider a circularly polarized pulse
to ensure the dependence of $a^2$ on $X=x-v_g t$
and existence of motion integrals (\ref{integrals}),
thus avoiding complication
that the laser pulse electromagnetic wave phase velocity
is equal to $v_{\rm ph,las} = 1/v_g > 1$.
We choose the finite quasi-Gaussian pulse shape
to emphasize the existence of the ``ponderomotive'' separatrix
(see below).
The pulse length is taken
to be less than, but not too much, the (relativistic) wakefield wavelegth
so as to make the wakefield excitation efficient enough
to preserve the effect of the laser field on the particle motion
inside the first period of the wakefield
and to simplify formulae below.

Each orbit $\{X(t),p_x(t)\}$ of the electron
in the $(X,p_x)$-plane is a segment of a level curve of the function $h(X,p_x)$.
The $(X,p_x)$-plane
is divided into basins of a finite motion,
where the particle is trapped by the wakefield potential,
and two basins of infinite motion.
The basins are separated from each other
by special orbits, separatrices,
which join at singular points
situated on the curve
$p_{xs}(X) = \beta_{\rm ph} \gamma_{\rm ph} \sqrt{1+a^2(X)}$.
On this line the square root in the right hand side
of Eq. (\ref{energy}) vanishes: 
$(\Delta\Phi + \chi_0)^2 - \gamma_{\rm ph}^{-2} ( 1+a^2(X)) = 0$.

An electron started from the singular point $X_s$
acquires the maximum kinetic energy
at the top $X_t$ of the separatrix,
as one can easily calculate from Eq. (\ref{energy})
substituting $X_0=X_s$, $X=X_t$,        
$p_{x0}=p_{xs}(X_s)$.
If the laser pulse length
is shorter than a half of wakefield wavelength,
then in the first period of the wakefield
the points $X_s$ and $X_t$ correspond, respectively,
to the local minimum and maximum of the wakefield potential,
$\Phi(X_s)=\Phi_{\rm min}$, $\Phi(X_t)=\Phi_{\rm max}$.
So the maximum kinetic energy on the separatrix is
\begin{equation}
{\cal E}_{\rm m} =
\gamma_{\rm ph}^2
\bigr(
  \Delta\Phi_{\rm m} +
  \beta_{\rm ph}
\big[ \Delta\Phi_{\rm m}^2 + 2\gamma_{\rm ph}^{-1}\Delta\Phi_{\rm m} \big]^{1/2}
\bigr)
+ {\cal E}_{\rm inj} 
\label{energy-max}
\, ,
\end{equation}
where $\Delta\Phi_{\rm m}=\Phi_{\rm max}-\Phi_{\rm min}$,
${\cal E}_{\rm inj} = \gamma_{\rm ph} - 1$.
If the laser pulse length
is much shorter than the wakefield wavelength
and $\gamma_{\rm ph} \gg 1$,
we have
${\cal E}_{\rm m} \approx
2 \gamma_{\rm ph}^2 \Delta\Phi_{\rm m} +
\gamma_{\rm ph} - 1$.
The lowest value of the potential $\Phi$
is reached when
the laser pulse sweeps the greatest possible amount of electrons
(in 1D -- a half of all the electrons per wakefield period),
$\Phi_{\rm min}\geqslant -1+1/\gamma_{\rm ph}$,
and the highest value is limited by the ion responce,
$\Phi_{\rm max}\leqslant \mu(1-1/\gamma_{\rm ph})$.
Knowing the minimum of the solution to Eq. (\ref{eq:phi}),
one can find its maximum; in the case
of a sufficiently short and intense laser pulse
($l_p\ll\lambda_{\rm wf}$, $a\gg 1$),
Eq. (\ref{eq:phi}) gives
$\Phi_{\rm max} = (1-\gamma_{\rm ph}^{-1})(2\gamma_{\rm ph}\mu+\mu-1)/(2\gamma_{\rm ph}+\mu-1)$,
which in the limit $m_i\gg m_e$ tends to $2\gamma_{\rm ph}-1-\gamma_{\rm ph}^{-1}$.
If the laser pulse has the optimal length,
then $\Phi_{\rm max} \approx a^2/2$ for $a\lesssim \sqrt{\mu}$,
\cite{Bulanov-a2}.


Since the laser pulse has a finite duration,
the ``runaway'' separatrix exists,
a segment of the level curve $h(X,p_x)=1/\gamma_{\rm ph}-\Phi_{\rm min}$,
Fig. \ref{fig:phase}(c).
If an electron beam is injected exactly
onto this separatrix,
it asymptotically {\it overtakes} the laser pulse
and becomes monoenergetic
with the final energy, as it follows from Eq. (\ref{energy-max}),
\begin{equation}
{\cal E}_{\rm f} =
\gamma_{\rm ph}^2
\!
\bigr(
  |\Phi_{\rm min}| +
  \beta_{\rm ph}
\big[ \Phi_{\rm min}^2 \!+ 2 \gamma_{\rm ph}^{-1} |\Phi_{\rm min}| \big]^{1/2}
\bigr)
+ {\cal E}_{\rm inj} 
\label{energy-fin}
\, ,
\end{equation}
where $|\Phi_{\rm min}|=-\Phi_{\rm min}>0$.
In the limit $\gamma_{\rm ph}\gg 1$,
this energy can be much higher than
the required minimum injection energy.
If, additionaly,
the wakefield is strongly nonlinear ($a \gg 1$),
$\Phi_{\rm min}$
tends to its lowest value $-1+1/\gamma_{\rm ph}$
and we have
${\cal E}_{\rm f, max}\approx
2 \gamma_{\rm ph}^2 + \gamma_{\rm ph} - 1$.


In the first period of the wakefield
behind the laser pulse
there is also the ``confined'' separatrix,
a segment of the level curve $h(X,p_x) \approx 1/\gamma_{\rm ph}$
(the exact value is discussed below).
It encloses a basin of orbits of electrons
which are trapped inside the potential well
moving along with the laser pulse.
Between the ``confined'' and ``runaway'' separatrices
lies a bunch of reflecting orbits.
On such an orbit
an electron starts with the longitudinal momentum $p_{x}^{-}$ in the range
$\beta_{\rm ph}\gamma_{\rm ph} >
    p_{x}^{-} >
	\beta_{\rm ph} \gamma_{\rm ph} +
	\gamma_{\rm ph}^2
	\bigl(
	\beta_{\rm ph} |\Phi_{\rm min}| -
	[ \Phi_{\rm min}^2 \! + 2\gamma_{\rm ph}^{-1}|\Phi_{\rm min}|]^{1/2}
	\bigr)
	\geqslant 0$
at $t\rightarrow -\infty$.
Then it is accelerated by the first cycle of the wakefield,
reaching the maximum energy defined by Eq. (\ref{energy}),
where one must substitute
$X_0=+\infty$, $p_{x0}=p_{x}^{-}, a(X_0)=\Phi(X_0)=0$.
Finally, the electron overtakes the laser pulse.
Its longitudinal momentum $p_{x}$
and kinetic energy ${\cal E}$ increase as
\begin{eqnarray}
p_{x}^{+} &=& p_{x}^{-} + 2\gamma_{\rm ph}^2 {\Gamma}^{-}(\beta_{\rm ph}-v_{x}^{-})
\, ,
\label{p-plus}
\\
{\cal E}^{+} &=& {\cal E}^{-} +
2\beta_{\rm ph} \gamma_{\rm ph}^2 {\Gamma}^{-}(\beta_{\rm ph}-v_{x}^{-})
< {\cal E}_{\rm f}
\, ,
\label{E-plus}
\end{eqnarray}
where
$\Gamma^{-} = [1+(p_{x}^{-})^2]^{1/2}$,
${\cal E}^{-} = \Gamma^{-} - 1$,
$v_{x}^{-} = p_{x}^{-}/\Gamma^{-} < \beta_{\rm ph}$.
The same equations describe an elastic rebound
of a relativistic particle 
from the wall moving
at a speed
$\beta_{\rm ph}$.


Yet another, the third, ``ponderomotive'' separatrix exists
in the vicinity of the laser pulse front $X_f=l_p$,
Fig. \ref{fig:phase}(c), (e).
It joins the second, ``confined'', separatrix
at the point $(X_{\rm p},p_{xs}(X_{\rm p}))$ defined by equation
$a(X_{\rm p})a'(X_{\rm p}) = \gamma_{\rm ph}\Phi'(X_{\rm p})\sqrt{1+a^2(X_{\rm p})}$
and so the exact value of the Hamiltonian for both separatrices
is $h_{\rm p}=h(X_{\rm p},p_{xs}(X_{\rm p}))$.
The third separatrix encloses a thin basin of
orbits with $1/\gamma_{\rm ph} < h(X, p_x) < h_{\rm p}=h(X_{\rm p},p_{xs}(X_{\rm p}))$,
going from  $X=+\infty$ at $t\rightarrow -\infty$ with $p_x<\beta_{\rm ph}\gamma_{\rm ph}$
and reflecting back 
with increased $p_x>\beta_{\rm ph}\gamma_{\rm ph}$
at $t\rightarrow +\infty$.
In contrast to orbits between the ``confined'' and ``runaway'' separatrices,
where particles are reflected by the wakefield potential,
the orbits enclosed by the third separatrix
belong to electrons
which are reflected (accelerated) by the {\it ponderomotive force} of the laser pulse.
Such reflection is possible because the laser pulse
has the speed $v_g < 1$
and the wakefield potential $\Phi(X)$
always grows slower than the $a(X)$ on the laser pulse front.
%
%
Using series expansions of functions $a(X)$ and $\Phi(X)$
about the point $X=X_f$,
$a(X_f  + \xi)  = a_1 \xi + a_2 \xi^2 /2 + o(\xi^3)$,
$\Phi(X_f  + \xi)  = k_p^2 \beta_{\rm ph}^{-2} (1 - \mu^{-2})
\left( a_1^2 \xi^4 /24 + a_1 a_2 \xi^5 /40 \right) + o(\xi^6)$,
where $a_1=a'(X_f)$, $a_2=a''(X_f)$,
we can estimate
the ``ponderomotive'' basin thickness,
which is the energy difference between
the upper and lower branches of the ``ponderomotive'' separatrix
\begin{eqnarray}
{\cal E}_{\rm p}^{+} - {\cal E}_{\rm p}^{-} =
2\beta_{\rm ph}\gamma_{\rm ph}^2 \sqrt{h_p^2-\gamma_{\rm ph}^{-2}}
\approx
\nonumber
\\
-\frac{2\sqrt{3}\, \beta_{\rm ph}}{k_p} a'(X_f) \gamma_{\rm ph}^{1/2}
+ \frac{21\sqrt{2}\, \beta_{\rm ph}^2}{5 k_p^2} a''(X_f)
\end{eqnarray}
at 
$|a'(X_f)|\ll 1$,
$|a''(X_f)|\ll 1$,
and $\mu \gg 1$, $\gamma_{\rm ph}\gg 1$.
The ponderomotive acceleration affects only those electrons
that move in the same direction as the laser pulse and
whose velocity is slightly less than $\beta_{\rm ph}$.
The acceleration gain turns out to be rather small,
because the ponderomotive and electrostatic potentials
almost completely compensate each other. 
However, it is still not zero even with the ideal Gaussian pulse;
the maximum effect is reached when the laser pulse
has a sharp front.

We examine the energy spectrum change
of an electron bunch
injected into the first period of the wakefield wave
onto the ``runaway'' separatrix, Fig. \ref{fig:phase}(c).
When a relatively long, initially quasi-mono\-energetic, bunch
is injected from the singular point $X_s$
and accelerated in the first period of the wakefield wave,
its particles are distributed
along the ``runaway'' separatrix with some density ${\cal N}(X)$.
As a result,
the particle energy spectrum broadens
from the initial energy ${\cal E}_{\rm inj} = \gamma_{\rm ph}-1$
to the cut-off (maximum) energy ${\cal E}_m$.
Besides these two limits,
the spectrum can have a peculiarity at ${\cal E}_{\rm f}$, Eq. (\ref{energy-fin}).
Near the top of the separatrix
the particle energy has a parabolic dependence on $X$,
${\cal E}(X) \simeq {\cal E}_m (1 - (X-X_t)^2/\Lambda^2)$,
where
$\Lambda^2 = -2{\cal E}_m/{\cal E}''(X_t)$,
as it follows from Eqs. (\ref{energy}), (\ref{eq:phi}).
Hence the energy spectrum near the cut-off energy is
\begin{equation}
\frac{dN}{d{\cal E}}
= \frac{{\cal N}(X)}{|d{\cal E}/dX|}
\simeq
\frac{\Lambda {\cal N}(X_t)}{2\sqrt{ {\cal E}_m ({\cal E}_m-{\cal E}) }}
\, ,
\label{distr-ne}
\end{equation}
where ${\cal E} < {\cal E}_m$.
Assuming that the density ${\cal N}(X)$ of the particle
distribution along the separatrix is smooth enough at $X=X_t$,
we see that the spectrum has an integrable singularity.
If the particles are arranged uniformly on the separatrix,
the spectrum, despite its singularity,
has rather large spread,
e.~g. a half of the particles
occupies the energy interval
$3{\cal E}_m/4 \leqslant {\cal E} < {\cal E}_m$.

\begin{figure}
\includegraphics{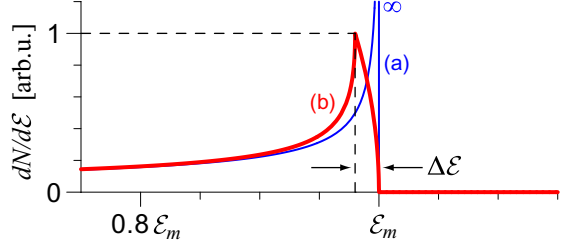}
\caption{\label{fig:spectra}
(color).
The energy spectrum of the electron bunch
scattered about the top $X_t$ of the separatrix
for
the bunch initial energy spread $\Delta{\cal E}=0$ (a)
and for $\Delta{\cal E} = 0.02 {\cal E}_m$ (b).
The distribution $dN/d{\cal E}$
unit is
$\frac{3N_b}{2{\cal E}_m}
  \frac{(\Delta{\cal E}/{\cal E}_m)^{1/2}}{1-(1-\Delta{\cal E}/{\cal E}_m)^{3/2}}$.
}
\end{figure}

In general, the bunch has some initial energy spread $\Delta{\cal E}$.
During acceleration
it occupies some region around the separatrix
in the plane $(X,{\cal E})$.
To estimate the bunch spectrum,
we assume that the particles are distributed uniformly
between two orbits which we approximate
by parabolas
${\cal E}_1(X) = {\cal E}_m-\Delta{\cal E} - {\cal E}_m X^2/\Lambda^2$
and 
${\cal E}_2(X) = {\cal E}_m - {\cal E}_m X^2/\Lambda^2$.
Correspondingly, the particle distribution function is
\begin{equation} \label{distr-xe}
f(X,{\cal E}) \! = \! \frac{
3 N_b\,
\theta[ X \! +\! \Lambda ]
\,
\theta[ \Lambda \! -\! X ]
\,
\theta[ {\cal E} \! -\! {\cal E}_1 ]
\,
\theta[ {\cal E}_2 \! -\! {\cal E} ]
}%
{
4\Lambda {\cal E}_m \;
\big( 1 - (1\! -\! \Delta{\cal E}/{\cal E}_m)^{3/2}
\big)
}
\, ,
\end{equation}
where $\theta$ is the Heaviside step function
($\theta[\xi]=1$ for $\xi\geqslant 0$ and $=0$ for $\xi<0$),
$N_b$ is the number of particles in the bunch.
Then the energy spectrum is
\begin{equation}
\!
\frac{dN}{d{\cal E}} \! =
\!\!\!
\int\limits_{-\Lambda}^{\Lambda} \!\!\! f(X,{\cal E}) dX \! = \!
\frac{3 N_b}{2}
\frac{
\sqrt{\smash[b]{{\cal E}_m \! -\! {\cal E}}}
\! - \!
\sqrt{\smash[b]{{\cal E}_m \! -\! \Delta{\cal E} \! -\! {\cal E}}}
}%
{
{\cal E}_m^{3/2} - \left({\cal E}_m\! -\! \Delta{\cal E}\right)^{3/2}
}
,
\label{distr-e}
\end{equation}
where the square root must be set to zero if its argument is negative.
In the limit $\Delta{\cal E}\rightarrow 0$
this spectrum tends to
$dN/d{\cal E}=
N_b \, \theta({\cal E}_m \! -\! {\cal E})/2\sqrt{ {\cal E}_m ({\cal E}_m-{\cal E}) }$,
which is equivalent to Eq. (\ref{distr-ne}) at
$N_b = \Lambda {\cal N}(X_t)$.
The sum of the spectrum (\ref{distr-ne})
over the range of ${\cal E}_m$
varying from ${\cal E}_{m} - \Delta{\cal E}$ to ${\cal E}_{m}$
weighted with $\sqrt{{\cal E}_m}$
yields Eq. (\ref{distr-e}).
Both the spectra (\ref{distr-ne}) and (\ref{distr-e})
are shown in Fig. \ref{fig:spectra}.

%
%
The above-stated analysis concerns
the phase portrait of a negatively charged particle
(electron).
In this paragraph we make digression
considering the case of a positively charged
particle, positron.
In this case formulae (\ref{h-hamilt})-(\ref{energy}) remain valid
with the substitution $\Phi \rightarrow -\Phi$.
The phase portrait of the positron in the same wakefield
and laser field as above is shown in Fig. \ref{fig:phase}(d).
In the wakefield,
the electron's points of equilibrium
correspond to the
positron's singular points
(for sufficiently short laser pulse).
The positron injected from the singular point
into the second cycle of the wakefield returns back
to the same singular point.
In the first half-cycle of the wakefield,
in contrast to the case of the electron,
both forces acting on the positron
--
the wakefield electrostatic force
and the laser pulse ponderomotive force
--
pull the positron in the same direction (``forward'').
Therefore we see a wide ``ponderomotive'' basin
in which the orbits with initial momentum
$\beta_{\rm ph}\gamma_{\rm ph} > p_{x,\rm pos}^{-} >
  \gamma_{\rm ph}^2
  \big( \beta_{\rm ph} \big(\Phi_{\rm p,pos}+\gamma_{\rm ph}^{-1} \sqrt{\smash[b]{1+a_{\rm p,pos}^2}}\big)
	-
        \big[ \big(\Phi_{\rm p,pos}+\gamma_{\rm ph}^{-1} \sqrt{\smash[b]{1+a_{\rm p,pos}^2}}\big)^2 - \gamma_{\rm ph}^{-2}
	\big]^{1/2}
  \big)
$
are accelerated
up to the energy
\begin{eqnarray}
{\cal E}_{\rm p,pos}^{+} =
 \gamma_{\rm ph}^2
 \Bigl\{ \Phi_{\rm p,pos} + \gamma_{\rm ph}^{-1} \sqrt{{1+a_{\rm p,pos}^2}}
\Bigr.
\nonumber
\\
\Bigl.
 + \beta_{\rm ph}
   \Bigl[ \Bigl( \Phi_{\rm p,pos}+\gamma_{\rm ph}^{-1} \sqrt{{1+a_{\rm p,pos}^2}}\Bigr)^2 - \gamma_{\rm ph}^{-2}
   \Bigr]^{1/2}
\Bigr\}
\, ,
\label{energy-ponderomotive-positron}
\end{eqnarray}
in accordance with Eqs. (\ref{energy}), (\ref{E-plus}).
Here the limiting values of the positron momentum and energy
in the ``ponderomotive'' basin are defined via
the wakefield potential $\Phi$ and the laser amplitude $a$
taken at the singular point $X_{\rm p,pos}$,
a non-trivial solution to the equation
%
%
\begin{equation}
a(X)a'(X) + \gamma_{\rm ph}\Phi'(X)\sqrt{1+a^2(X)} =0 \, .
\end{equation}
The momentum of the lower branch of the ``ponderomotive'' separatrix is negative at
$\Phi_{\rm p,pos}+\gamma_{\rm ph}^{-1} \sqrt{\smash[b]{1+a_{\rm p,pos}^2}} > 1$,
then the positron initially at rest is accelerated
up to momentum $2\beta_{\rm ph}\gamma_{\rm ph}^2$
and energy $2\beta_{\rm ph}^2\gamma_{\rm ph}^2$,
in accordance with Eqs. (\ref{p-plus}), (\ref{E-plus}).
Thus even the ``background'' positrons,
introduced externally or created in the laser-plasma interaction,
are substantially accelerated.
In the limit of a long laser pulse ($l_p \gg \lambda_{\rm wf}$),
the maximum energy (\ref{energy-ponderomotive-positron})
becomes $\approx (1+\beta_{\rm ph}) \gamma_{\rm ph}^2 a_0$,
since $\Phi_{\rm p,pos}\approx a_0$
in this limit.
%

In conclusion,
in the first cycle of the Langmuir wave in the wake of
the short relativistically strong laser pulse
electrons have at least three separatrices:
on the ``runaway'' separatrix the electron overtakes the wakefield
and the laser pulse,
on the ``confined'' separatrix it moves together with the laser pulse
and the ``ponderomotive'' separatrix places quite tight limit
for the ponderomotive acceleration.
In contrast to electrons, positrons see the
wakefield and the laser pulse ponderomotive force acting in the same direction.
It is shown that
the energy spectrum of the initially mono-energetic particle bunch
spread about the top of the separatrix 
has a typical localized shape $\propto ({\cal E}_m-{\cal E})^{-1/2}$
with a sharp cut-off.

We thank
the Ministry of Education, Culture, Sports, Science and Technology of Japan
and CREST, Japan Science and Technology Agency for their support.


\end{document}